\documentclass[review]{elsarticle}

\usepackage{multirow}
\usepackage{multicol}

\usepackage[final]{changes}
\usepackage{url}
\usepackage[english]{babel}
\usepackage[T1]{fontenc}
\usepackage[utf8]{inputenc}
\usepackage{balance}

\usepackage{graphicx}

\DeclareGraphicsExtensions{.pdf,.jpg,.png}

\begin{document}

\title{D-LITe: Building Internet of Things Choreographies}
\author[mlv]{Sylvain Cherrier}
\author[lr]{Yacine M. Ghamri-Doudane}
\author[mlv]{Stéphane Lohier }
\author[mlv]{Gilles Roussel }
\address[mlv]{Université Paris-Est Marne-la-vallee (UPEM), Laboratoire d'Informatique Gaspard-Monge (LIGM) CNRS UMR 8049
	, 77454 Marne-la-Vallee Cedex 2,\\Email: fistname.lastname@u-pem.fr}
\address[lr]{L3iLab, Université de La Rochelle, Av. Michel Crepeau, 17042, La Rochelle CEDEX 1, France.,\\Email: yacine.ghamri@univ-lr.fr}

\begin{abstract}
In this work, we present a complete architecture for designing Internet of Things applications. While a main issue in this domain is the heterogeneity of Objects hardware, networks and protocols, we propose D-LITe, a solution to hide this wide range of low layer technologies. By abstracting the hardware, we focus on object's features and not on its real characteristics. D-LITe aims to give a universal access to object's internal processing and computational power. A small virtual machine embedded in each object gives a universal view of its functionalities. Each object's features are discovered and programmed through the network, without any physical access. D-LITe comes with the SALT language that describes the logical behaviour needed to include user's Objects into an IoT application. This communication is based on REST architecture. Gathering all these logical units into a global composition is our way to build a services Choreography, in which each Object has its own task to achieve. This paper presents also an analysis of the gain obtained when a Choreography is used instead of the most common services Orchestration.
\end{abstract}

\maketitle

\section{Introduction}
As Internet of Things becomes an increasingly attractive domain, there are still issues that limit its rise. Because of the diversity that came from the fundamental idea of IoT (interconnecting the well-known Internet to the user's wireless networks, provided by smart phones, home automation, smart buildings and smart cities), Things may encounters real difficulties to interact. If a standardized network protocol for all (IPv6) seems to arise while ensuring this communication, it appears that the development of applications involving all stakeholders are difficult to achieve. Things hardware has varied architectures, characteristics and processing power. Programming requires a certain level of skill and knowledge to be able to build an application involving such various elements. Demonstration and implementation of IoT applications are often confined to simple devices remote control through the Internet, using a smartphone for example. This is not what we expected from this field, that should tend to offer real and transparent self interactions between objects, without human intervention. This pervasive computing has been predicted by M.~Weiser in his visionary paper~\cite{WeiComputer21century1999}. Our goal is to provided a solution that gives Objects an easier way to be interconnected and to interact as part of a global application.

To reach that goal, we have analysed IoT applications under several angles. First, we have defined the framework in which we want to work. The IoT involves a wide range of components, so we chose to concentrate on the most constrained ones. The construction of a global set such as the IoT must take into account the constraints of the smallest participant, under penalty of exclusion. In our case, we have focus on the WSAN (\textit{Wireless Sensor and Actuator Network}). Our aim is to provide a solution running over sensors and actuators which are very constrained in term of energy and processing power. More powerful hardware will therefore raise no problem.

Our main idea is to hide each Object behind its logical representation. The more abstract the representation is, the more universal the element becomes. If we can perceive each object as a universal element, its involvement in an application, its interactions with others, and even its substitution by a different one (offering the same kind of features) become feasible. Reaching this universality is achieved through the use of an \textit{Hardware Abstraction Layer} (HAL)  hiding complex and specific API calls. This HAL is manipulated through a simple programming language (that can be understood by every supported Objects) through REST methods (a common and well-know architecture for the Internet) to access each node involved. REST architecture has been chosen because of its lightness. As every Object becomes remotely accessible and programmable, our solution builds a Services Choreography, and exempts the need for specific programming.

A Choreography is an architecture of distributed logical entities, working on their own. Each element interacts, and the application has no central point of control. The composition of all behaviours achieves the desired application. The choice between a Choreography or an Orchestration is made at the application level. In the Internet, this choice is driven by software or organization needs. But in the IoT, and more precisely in the WSAN (part of the IoT), this choice may have consequences regarding the constraints of this specific network. So we have compared the impact of the two alternatives in terms of energy and reliability in a network organized as a tree, a typical WSAN infrastructure .

This paper is organized as follows: Section II discusses the background and visions for IoT programming approach. Section III is an analysis of the impact of choreographed architectures over network organized as a tree. As WSAN is an important actor of the IoT, the specific constraints of this network promote a particular organization out of the two possible. This section quantifies the gain a Choreography can offer compared to an Orchestration~\cite{cherrier2012Quantify}. Section IV defines the major features or our platform D-LITe (\textit{Distributed Logic for Internet of Things sErvices})~\cite{cherrier2011Dlite} providing the hardware abstraction and the REST access to Objects. Section V describes SALT (\textit{Simple Application Logic Description using Transducers for Internet of Things})~\cite{cherrier2013SALT}, the language designed to express in a universal way the logic to run on each Object. Finally, Section VI concludes the paper.

\section{Background and vision}
The Internet of Things has several definitions. The basic idea is to allow communications between everyday-Objects (from their specific wireless communication capabilities) and everyday-services (from the widespread Internet) in a pervasive way, in order to build new usages and applications. Depending on their domain and knowledge, every actor of the domain has his own definition of the IoT~\cite{AtzoriIoT2010}. 

\subsection{What are the Objects of the IoT?}
Because Internet of Things involves a large aggregation of hardwares and networks, there are various approaches about it\cite{AtzoriIoT2010}. One of the IoT specificities is the collaboration of new networks with the global Internet. From Smart Phones to Personal Area Network(PAN) or Wireless Sensors and Actuators Networks (WSAN), more and more everyday-life objects get computational capabilities and wireless connectivity\cite{wang2013research}. As the communication with tools increases, the idea of interconnecting all these elements leads to a new architecture, mixing the Internet (of Data) and these new tools (with physicals capabilities).

But mixing such heterogeneous elements to build a unique domain arises new issues. Some elements have big computing capabilities, large memory and unlimited energy while others are very constrained. Some can interact with the real world, sensing events, measuring physical values, or acting on the physical world, while others are just virtual services. Some are mobiles, public or private, shared between users or only accessible to a unique user\cite{sundmaeker2010vision}.

Dealing with all differences can be done in different ways. For example, a gateway can adapt the communication by connecting two worlds. Another approach is to design a universal view of \textit{what is an Object}, by the definition of users needs, depending on the capabilities Objects have in common\cite{karagiannis2015survey}.

Our approach is rather to be classified in the second category: we have identified requirements of IoT applications. Then, we have matched them with what is common in the variety of elements that constitutes the IoT.

\subsection{Object-as-a-Service}
To get rid of the constraint of Objects diversity, we propose a vision of each element that frees us from their real implementation. The common solution is to access the values sensed by object or to call actuations provided by it, following the description furnished by a standard (oneM2M, AllJoyn, etc)~\cite{gubbi2013internet}~\cite{datta2015onem2m}~\cite{yun2015demo}~\cite{al2015internet}. We propose a top-down approach in which we access the computing power of the Object, no matter its specificities. A service running on each of them gives a universal access to the CPU of each object, and then to its internal sensors or actuators. Our \textit{ Object-as-a-service} point of view focuses on remotely use object's processing ability through an interface~\cite{cherrier2014object}. The communication with an element through an abstraction belongs to the Services Oriented Architecture (SOA) realm~\cite{erl2005service}. The idea of using such paradigm for Sensor Network has already been presented in TinySOA~\cite{rezgui2007service}. \textit{Object-as-a-service} gives a unique vision of Objects (their processing power) for the different stakeholders involved in the IoT, and allows more flexibility in the composition of real IoT applications. However, the main SOA protocol (SOAP) is not adapted to constrained objects and network because of its verbosity. REST architecture~\cite{RESTFielding} is an alternative to SOAP for distributed applications. REST is lightweight and does not need an XML Parser that may override constrained objects capacities. It simply uses HTTP methods to give access to the characteristics with the GET, PUT, POST and DELETE standard commands. Our proposal is to use these methods to send a description of the behaviour of this object (a program) in the global application. Instead of accessing to the data themselves, we propose to use REST to give an access to the processing of the data gathered by the object (or to actions on the real world through its actuators).

\subsection{Services: Orchestration or Choreography?}
In Services Oriented Architecture, two main organizations are used. A centralized one, called \textit{Orchestration}, and a distributed one, called \textit{Choreography}~\cite{peltz2003web,barros2006standards}. In Orchestrations, a central point has the control of the application workflow. This Orchestrator organizes the calls to the different services, gets the result, and has a global and complete view of the logic. On the other hand, a Choreography has no central control point, and services interact with others following their own logic. Each node knows what to do, and reacts to context's changes. In a choreography, a node is like a dancing couple in a ballroom. Each couple knows its steps, and reacts on events of the very near environment. There is no centralized control of any supervisor; decisions are mainly made at the closest level~\cite{Duhart2016}. 

This approach is not new in WSAN. Specific energy constraints has lead researchers to propose distributed solutions, making sensors directly interact with actuators, avoiding the unique point of control. I.F.~Akyildiz named it ``automated architecture''~\cite{akyildiz2004wireless} compared to the ``semi-automated'' mode that can be seen as an Orchestration. As it uses Choreographies, our solution delegates small parts of the global application to each participant, using processing capacities closer to the needs, saving bandwidth and therefore energy~\cite{mottola2010programming}.

\subsection{Our vision}


There is a main difference between traditional SOA as found in Internet of PC, and our IoT applications point of view. In traditional SOA, a software architect looks for the different services that are already running on multiple servers, and compose them in order to obtain the results desired. Traditional SOA uses pre-existing and running Web Services to focus on their composition. Usually, Web Services involved in SOA applications are developed using a general approach. In IoT applications, the behaviour of each element is not as rigid and so strictly pre-defined. It may depend on the context, on previous given reactions or on changing user's needs. In IoT, many Objects are used by a reduced number of users compared to the traditional SOA usage in the Internet of PC. Actually, \textit{one} user owned \textit{many} Objects, while \textit{a} traditional Web Services is used by \textit{many} Internet users~\cite{sundmaeker2010vision}.

We can build new applications by describing the needed services, create them, and then deploy them. The \textit{service} here is the processing of the data sensed by the object, or the actions it can do. The resulting application is a composition of distributed \textit{on-the-fly} made services, dynamically created following the user's needs. These services did not pre-exist. This ``on-demand'' services creation requires the ability to easily program and deploy new services. To be valid, our solution must provide a way to send the description of the specific logic of the desired service to the right node, taking into account object's functionalities, in a universal language. To be valid, our solution must cover the needs of IoT applications. It must also take into account all the constraints of every stakeholders.

The remainder of this paper describes our idea of a universal virtual machine, how to hide the real specific calls behind a Hardware Abstraction Layer, how to remotely program a device in an interoperable way adapted to the needs of IoT while dealing with the hard constraints of its components, in order to build applications that can cover an important part of user needs. 

\section{Choreography versus Orchestration}

\begin{figure} 
\centering 
\includegraphics[width=8cm]{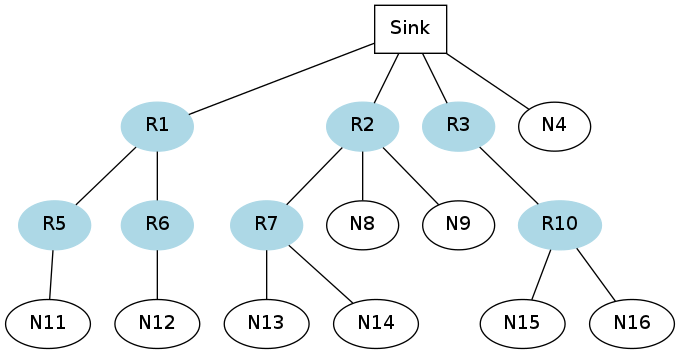} 
\caption{In this tree, the path length between node N9 and N13 is:\newline
1) in a Choreography : 3 hops (R2-R7-N13)\newline
2) in an Orchestration : 5 hops (R2-Sink-R2-R7-N13)}
\label{fig:BaseTree} 
\end{figure}
\begin{figure}[t] 
\centering 
\includegraphics[width=12cm]{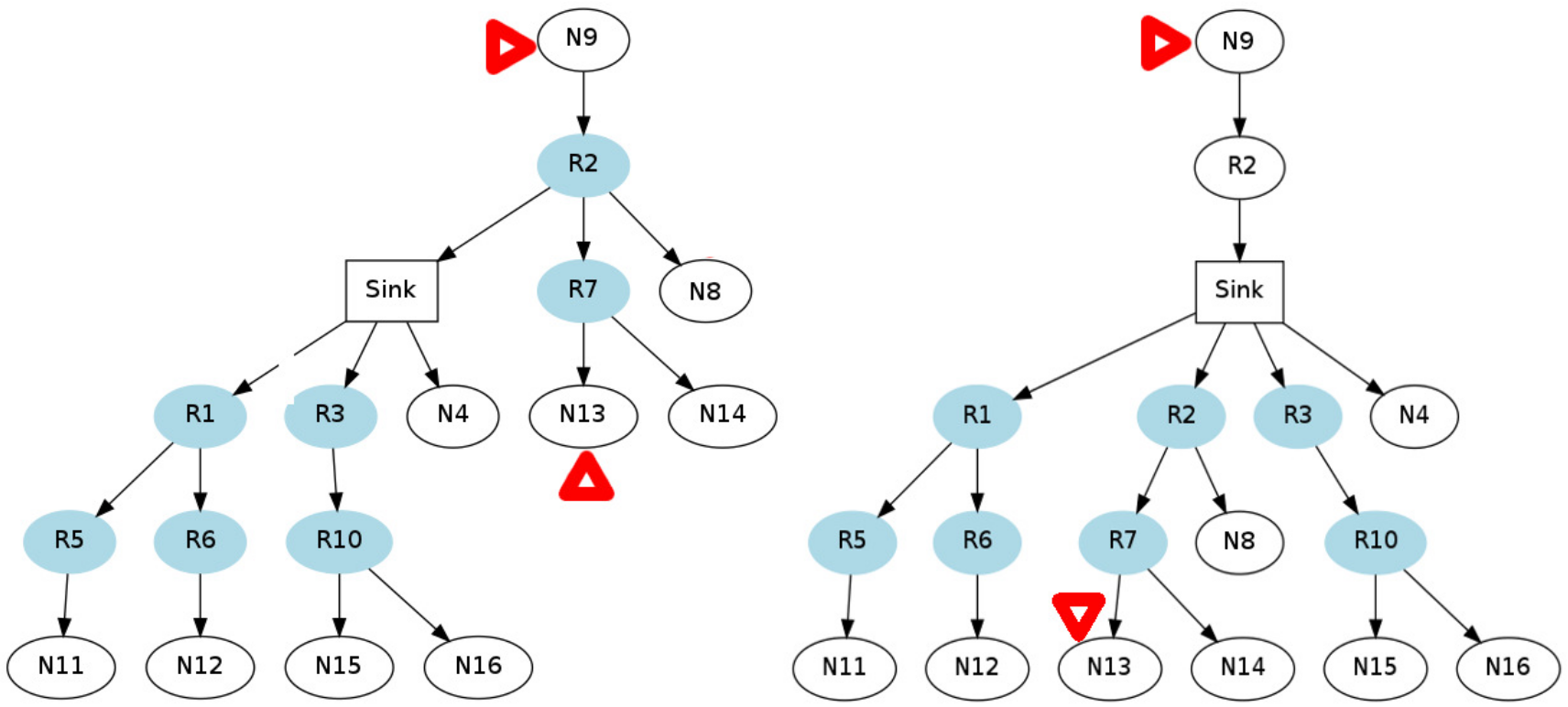}
\caption{From application perspective in a node, the tree is used to reach the destination. Here, the path from node 9 to node 13, from node 9 point-of-view in \textit{Choreography} (left) or \textit{Orchestration} (right)}
\label{fig:BaseChorOrch} 
\end{figure}

 
In this section, we are going to focus on the way the design of applications may impact some typical networks, and specifically WSAN. Tree architectures are often used in the case of WSAN organizations, such as Zigbee\cite{ZigBeeWebsite} or RPL\cite{shelby2010embedded} in 6LowPAN\footnote{ZigBee and 6LowPAN are two main network layer protocols for the IEEE standard 802.15.4}. Assuming our network is a tree (Fig~\ref{fig:BaseTree}), we have studied the impact of application's design, considering this application runs in a network organized as a tree. For example, the Figure~\ref{fig:BaseChorOrch} shows a WSAN of a 16 nodes. In this example, node 9 is a sensor, and node 13 an actuator. The user wants node 13 to react to events detected by node 9.

When the application is organized as an Orchestration, the messages from node 9 go to node 2, then to the sink (Fig~\ref{fig:BaseChorOrch}). The Orchestrator (often outside this network) computes the data. A new message, containing the resulting action, comes back from the Orchestrator, through the sink for node 2, then to node 7, and then to node 13, its final destination. The full path  (node 9/Orchestrator/node 13) is five hops long. Node 2 is used twice in this path.

A \textit{Choreographed} application is designed as a direct collaboration between nodes. The application, designed now without central decision point, distributes the logic as collaborations all over the network. The decision about the action to take is made directly on nodes that detect the event. For example, node 9 will decide to directly send an order to node 13. Based on this application's design, the node's view of the global network changes. Figure~\ref{fig:BaseChorOrch} shows the Choreographed path from node 13 to node 9, made of 3 hops. The message~\footnote{The message sent by node 13 in a Choreographed version is an order for node 9, while it was the sensed data in the Orchestrated version} is delivered more quickly, with less risk of getting lost. Node 2 has only one message to forward. Collision risks decrease. The single point of failure disappears.

The strong constraints (energy, memory, throughput) of WSAN are counterbalanced by the advantage of having processing power everywhere, even if it is very limited. The way these objects are used and interact is also a new domain. It takes advantage of the local processing power to relieve the entire network of useless messages~\cite{Duhart2016}. It results less energy consumption and lower radio traffic.

Our study is not limited to WSAN, but, as part of the IoT, sensors networks are the most constrained. We adopt the perspective of the developer, in a \textit{SOA} point-of-view. We do not evaluate here the appropriateness of the proposed network organization, but rather the consequences of the two application's design approaches on the existing organization. We consider our network as a tree of nodes. In such an organization, it seems that \textit{Choreography} uses some shorter ways in many cases (in facts, to all nodes that are on the path to the sink, or connected to that path, Fig~\ref{fig:BaseChorOrch}) as already presented by I.F.~Alkyildiz~\cite{akyildiz2004wireless} or L.~Mottola~\cite{mottola2010programming}. Our motivation here is the quantification of the improvement.

\subsection{Mathematical study}

We have studied the exchanges between nodes by calculating the length of all possible paths. Each path presented in Figure~\ref{fig:BaseTree} is considered following the two designs: \textit{Choreography} or \textit{Orchestration}. Depending on that software design, the vision changes for each node (Figure~\ref{fig:BaseChorOrch}).

\begin{figure}[t] 
\centering 
\includegraphics[width=8cm]{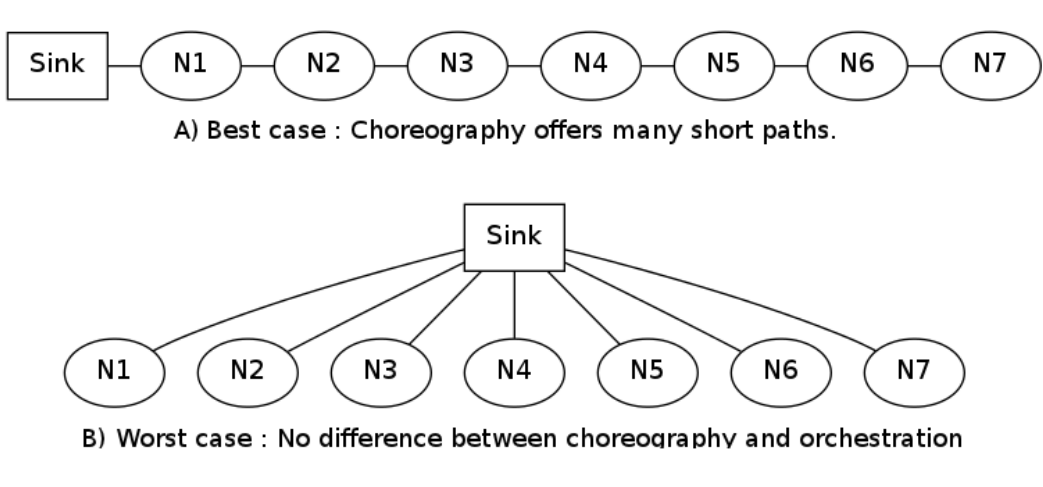} 
\caption{Extreme cases in \textit{Orchestration}/\textit{Choreography} comparison. The biggest difference is obtained in a inline-tree, the smallest (in facts, no difference at all) in a one-hop tree.} 
\label{fig:BestWorst} 
\end{figure} 

We have determined \textit{the best} and \textit{the worst} cases (Figure~\ref{fig:BestWorst}) for the two architectures.

In the \textbf{best case} (an in-line tree), nodes are numbered from the sink to the last node (0 to \emph{n}). We study the distance between nodes \emph{i} and \emph{j} (the path length). The formula giving the average path length for all nodes in Figure~\ref{fig:BestWorst}:A is :\newline

\[
 \mu_(n)=\frac{2}{n(n-1)} \sum_{i=1}^{n-1} \sum_{j=i+1}^n distance_{i,j}
\]
This formula can be simplified in
\[
 \mu_(n)= \frac{2}{|P|} \sum_{P} distance_{i,j}
\]
\[
\textrm{with } P = \textrm{ all couples in a set of } n \textrm{ nodes}
 \]
In an \textit{Orchestration}, data goes from node \emph{i} to the sink (node 0), and then back to \emph{j}. The distance between two nodes \emph{i} and \emph{j} is $(i+j)$.
\begin{equation}
\label{eq:Orchestration}
\mu_o(n)=\frac{2}{|P|} \sum_{P} (i+j) =n + 1
\end{equation}
Considering a \textit{Choreography}, the distance between \emph{i} and \emph{j} is shorter:~$(j-i)$.
\begin{equation}
\label{eq:Choreography}
\mu_c(n)=\frac{2}{|P|} \sum_{P} (j-i) = \frac{1}{3}(n + 1)
\end{equation}
After simplification, the average path length between two nodes in a linear structure of \emph{n} nodes is $n+1$ in the case of \textit{Orchestration}. and $1/3(n+1) $ for a \textit{Choreography}. In a linear tree, a \textit{Choreographed} architecture reduces the average path length by 3 compared to an \textit{Orchestration} (Figure~\ref{fig:BestWorst}(A)).

The \textbf{worst case} (Figure~\ref{fig:BestWorst} (B)) is a one-hop-tree where the path length between two nodes is always 2 hops. \textit{Choreography} and \textit{Orchestration} gives the same average path length. In such a tree, there is no difference between the two organizations. Finally, there is no case in which \textit{Choreography} has a longer path length than \textit{Orchestration}.

In conclusion, when using a \textit{Choreography}, the \textbf{best case} is encountered in a linear tree where each node has no sibling. Other trees (\textbf{general case}) will give advantage to \textit{Choreography} in a smaller ratio. Consequently, \textit{Choreography} always offers shorter average path length, except in the \textbf{worst case} scenario where both designs are strictly equivalent.

\subsection{Probabilistic study}

\begin{figure} 
\centering 
\includegraphics[width=12cm]{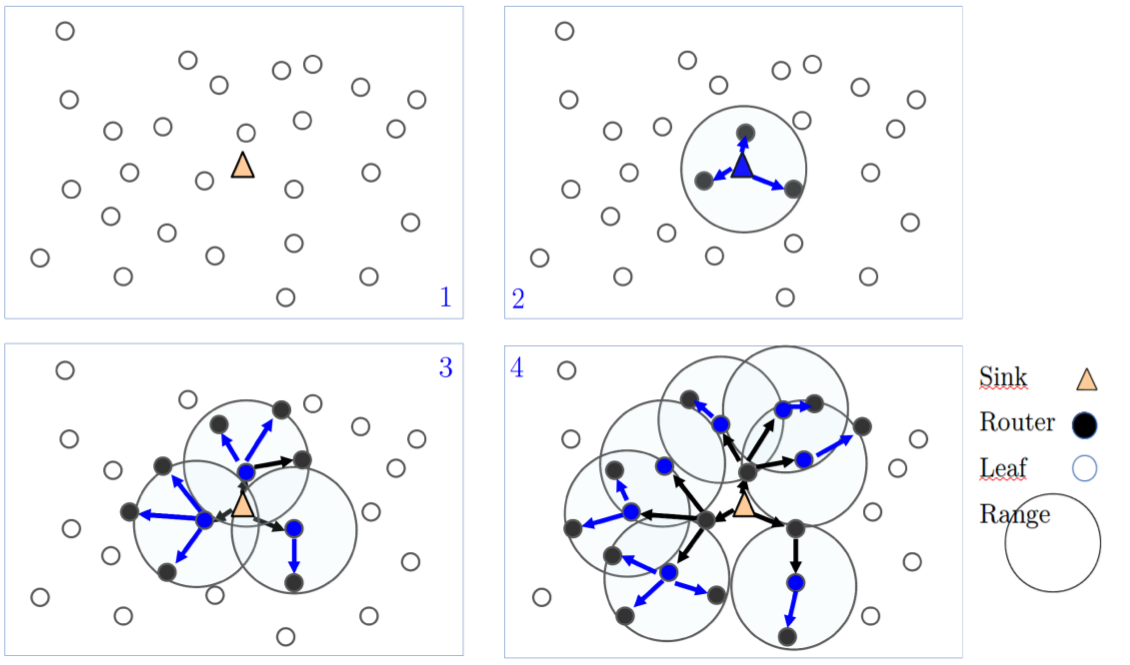} 
\caption{Algorithm used to build the tree. The sink (S) searches its neighbours (step 1 and 2),
which in turn look for their neighbours (step 3), and so on (step 4...). }
\label{fig:BaseModel} 
\end{figure} 
We have build a probabilistic model to study the \textbf{general case} (a common tree), by randomly positioning nodes on a square (Figure~\ref{fig:BaseModel}). The tree root (the sink) is at the center. A radius parameter defines the radio range for each node. The accessibility is estimated with a Unit Disk Graph (UDG). Any node beyond this radius is considered as inaccessible. At the beginning of our simulation, the sink starts to find reachable nodes. Then, these first level nodes try to reach other nodes, and so on (Figure~\ref{fig:BaseModel}). The resulted tree is analyzed for \textit{Orchestration} and \textit{Choreography} usages. For each design, the simulation gives the distribution of the number of paths by length.

\begin{figure} 
\centering 
\includegraphics[width=12cm]{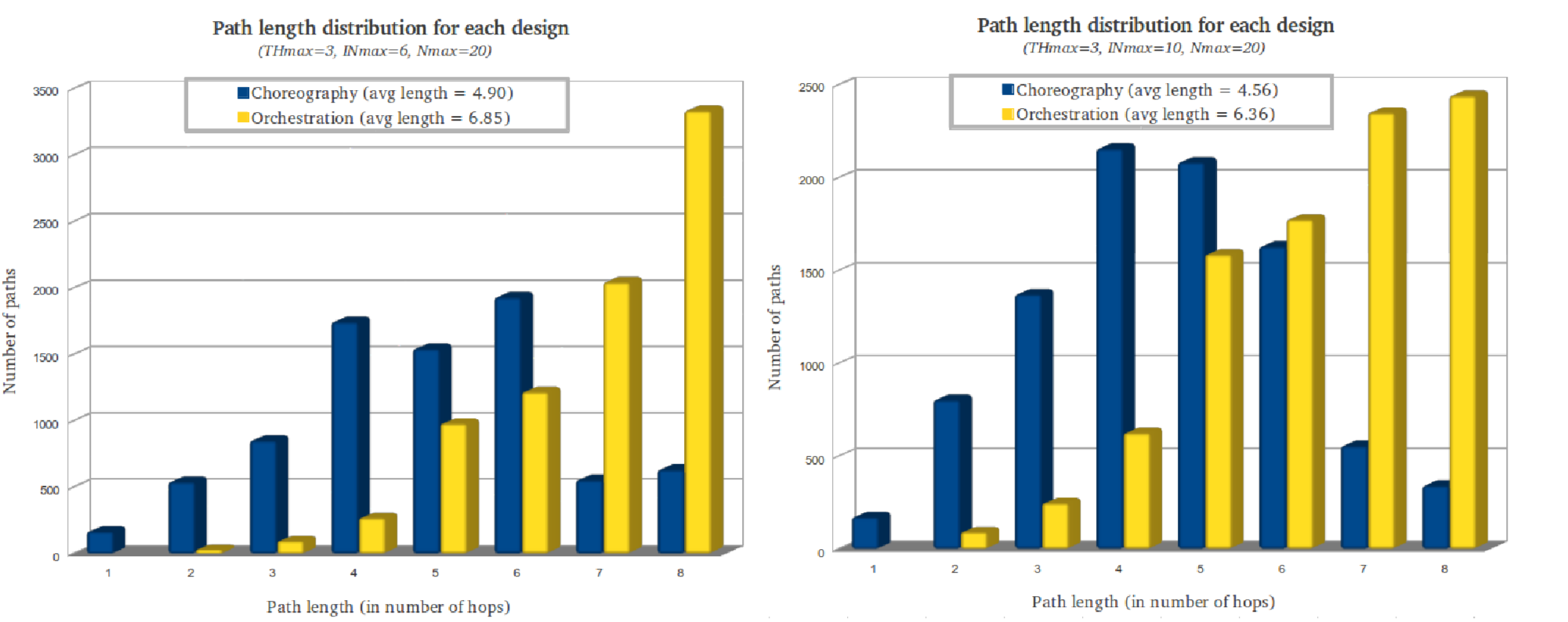}
\caption{Comparison between \textit{Choreography} and \textit{Orchestration} for a tree of reduced height(depth 3) and for usual values given in ZigBee presentation (left) or wider (right).} 
\label{fig:ZigBeeTree} 
\end{figure}

\begin{figure}[t] 
\centering 
\includegraphics[width=8cm]{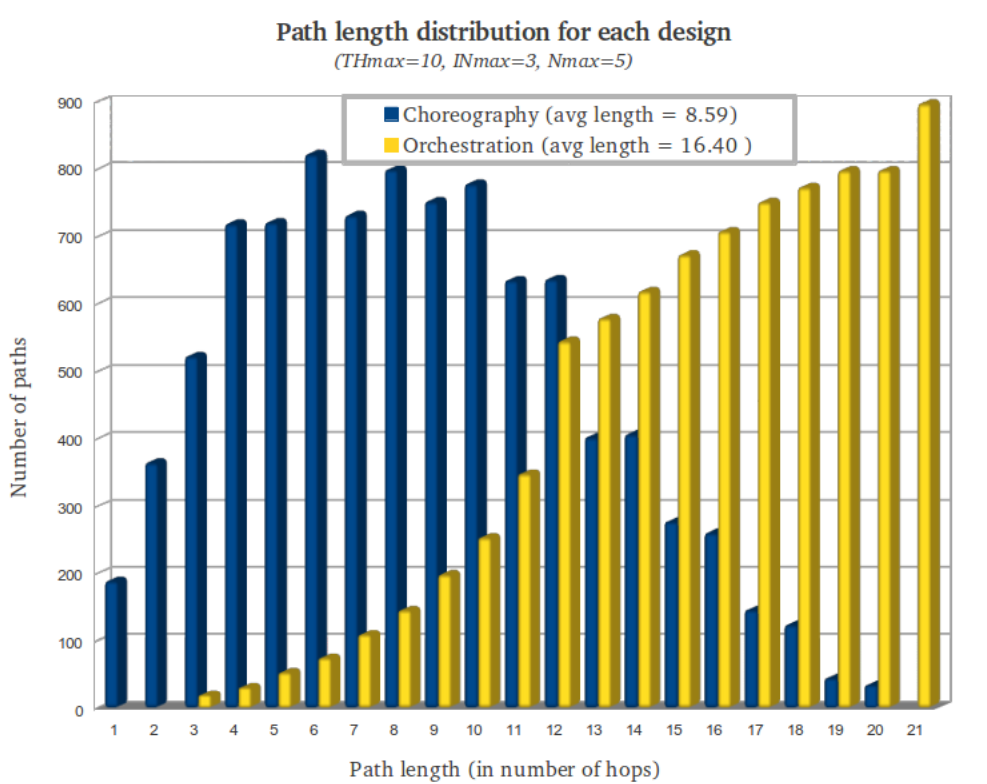}
\caption{A favorable case for Choreography.} 
\label{fig:ThinZigBeeTree} 
\end{figure}

The results in Figures~\ref{fig:ZigBeeTree} and~\ref{fig:ThinZigBeeTree} compare these two architectural designs, depending on the number of nodes and the three following parameters: \textit{THmax}  (Tree Maximum Height), \textit{INmax} (Internal Nodes Maximum number) and \textit{Nmax} (Nodes Maximum number). The variation of these parameters generates various forms of trees. For example, ZigBee~\cite{ZigBeeWebsite} uses the same kind of parameters \textit{Lmax, Rmax and Cmax} to build the network.

Our experiment involves 100 nodes. The three parameters (\textit{THmax, INmax and Nmax}) vary for each graph, in order to build the structure of each tree. We have calculated the length of each path from each node to all other nodes, according to \textit{Orchestrated} architecture (via the sink) and then to \textit{Choreographed} architecture (using the shortest path in the tree). 1000 tests were done in each analysis. The distribution of average path length of 10 analyses is plotted on Figures~\ref{fig:ZigBeeTree} and~\ref{fig:ThinZigBeeTree}.

The distribution of paths length for a tree with a maximum height of 3 with 20 children and 6 internal nodes maximum is given Figure~\ref{fig:ZigBeeTree} (on the right). In \textit{Choreography}, the paths length is generally and significantly shorter than in \textit{Orchestration}. The path length distribution for trees using the usual values of ZigBee~\cite{ZigBeeWebsite}) (\textit{THmax=3, INmax=6, Nmax=20}) is plotted on Figure~\ref{fig:ZigBeeTree}(left). This ``ZigBee demo'' tree is less compacted than the first one. The number of paths of maximum length increases in the \textit{Orchestration} case. In this configuration, the \textit{Choreography} is still a better choice in terms of path length.

Figure~\ref{fig:ThinZigBeeTree} is a favorable network configuration for \textit{Choreography}: important tree height, few internal nodes and leaves at each level. Here again, \textit{Choreography} is the more efficient organization in terms of path length.

\begin{table}[t]
\caption{Theoretical average path lengths according to the application design}.
\begin{tabular}{|l|r |r |r|}
\hline
\multirow{2}{*}{A \textit{n} nodes network} & \multicolumn{3}{|c|}{Average path
length}\\
\cline{2-4} &Orch. & Chor. & Ratio C./O. \\
\hline
Worst Case (THmax=0) & 2 & 2 & 100\% \\
\hline
THmax=3 INmax=3 Nmax=10 & 6.96 & 5.67 & 81 \% \\
\hline
 THmax=3 INmax=6 Nmax=20 & 6.85 & 4.90 & 72 \% \\
\hline
 THmax=3 INmax=10 Nmax=20 & 6.36 & 4.56 & 71 \% \\
\hline
 THmax=10 INmax=3 Nmax=5 & 16.40 & 8.59 & 52 \% \\
\hline
Best case (THmax=n)& (n+1) & 1/3(n+1) & 33 \% \\
\hline
\end{tabular}
\label{table:Gains} 
\end{table}

The result of our probabilistic study (Table~\ref{table:Gains}) complies with our two models: the ``best and worst case'' and the ``general case''. As the tree height increases, the \textit{Choreography} still gives shorter averaged path length. When there is a small number of internal nodes, a reduction by a factor 2 can be observed, as presented on Figure~\ref{fig:ThinZigBeeTree}.

The mathematical analysis shows than the \textit{Choreography} is always a better choice in terms of path length when using a network organized as a tree. The gain varies from 1 to 3 times compared to an \textit{Orchestrated} architecture. The improvement depends on the characteristics of the network topology (the form of the tree) and the number of nodes involved.

\subsection{Experimental study}
To verify our results on a real platform, we have experiment these two architectures on Contiki~\cite{dunkels2004contiki}\footnote{Contiki-OS is an Open Source Operating System for small Objects}. Contiki comes with an emulator to tests binaries, called Cooja. Experiments were made with Cooja.

\begin{figure}[t] 
\centering 
\includegraphics[width=8cm]{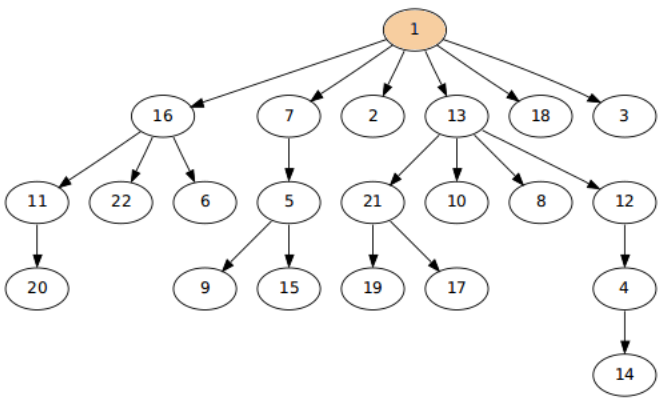} 
\caption{Tree given by Contiki during experiment \#50. High-level nodes are 16, 7 and 13.} 
\label{fig:Chor50} 
\end{figure} 

\begin{figure}[t] 
\centering 
\includegraphics[width=12cm]{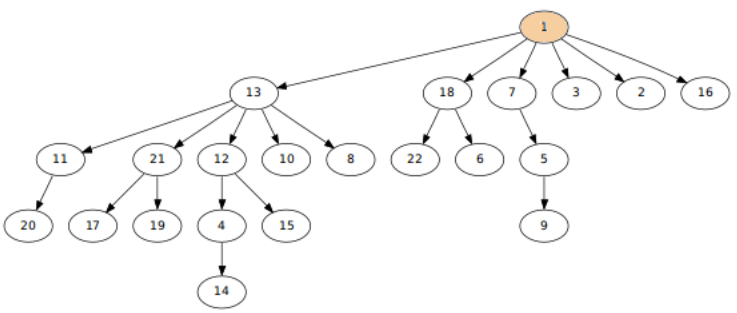} 
\caption{Experiment \#60: node 16 is no more a router. Node 18 is now in charge of node 22 and 6. 13 becomes more important.}
\label{fig:Chor60} 
\end{figure} 

Contiki is an Operating System for small objects  with a 802.15.4 transmission chipset. It offers an implementation of 6LowPan and RPL. 6LowPAN is an adapted version of IPv6 for WSAN~\cite{kushalnagar2007rfc} and RPL is the routing protocol~\cite{tsiftes2010low} in charge of building a tree for multi-hop communication inside the WSAN. The routing tree is automatically managed by RPL, and may be rebuilt dynamically if needed. We have gathered data from Contiki: the form of the tree, the number of \textit{sent}, \textit{received} and \textit{forwarded} IP packets, in order to monitor the global network activity. We have also collected the number of messages that have been successfully received by our application at the user level. In this paper, we will use the word \textit{packets} to mean the \textit{whole network activity}, and \textit{messages} in reference to the \textit{application} layer.

\subsection{Description of the experiment}
In our testbed, the program maker has no control on the shape of the RPL tree. Our experiment aims to test the impact of the application design over such a tree. It implements the following algorithm: each node chooses randomly one recipient to simulate an interaction between a sensor and an actuator. Then, each node sends 30 messages. We have counted the number of messages which currently reach the destination. The results (number of received messages and number of IP packets over the network) are achieved using both architectures: either by sending directly to the recipient (\textit{Choreography}) or through the sink (\textit{Orchestration}). The dynamic changes in the network (Fig~\ref{fig:Chor50} and~\ref{fig:Chor60}) and the randomized association between nodes give an overview of the effects of software architecture over the network.

22 Nodes are positioned randomly in Cooja, and node 1 acts as the sink. We have run 100 experiments for each architecture. Messages are transmitted through the sink in the case of \textit{Orchestration}, while they use the shortest path for \textit{Choreography}.

\subsection{Testbed's first result: nodes activity by level}
By using randomized association between a sensor and an actuator, and by regularly defining a new pair, we create the diversities that one could find in the real world. Indeed, in a real application, the link between 2 nodes is not automatically short. Nodes can be distant, and even if they are close, the network path between them may use many hops (the geographical position and the network topology does not automatically match). The first results obtained with this experiment are ignored because the load generated by the construction of the RPL tree may have distorted the results. To identify the major trends, the remaining measures are grouped into three classes corresponding to the different level of activity detected at the top level nodes (acting as routers, i.e. nodes 16, 7 and 3 in Figure~\ref{fig:Chor50}). Classes are: \textit{low}, \textit{average} and \textit{high} activities. For each class, we have evaluated the PDR (Packet Delivery Rate) of the whole network.

\begin{figure}[t] 
\centering 
\includegraphics[width=12cm]{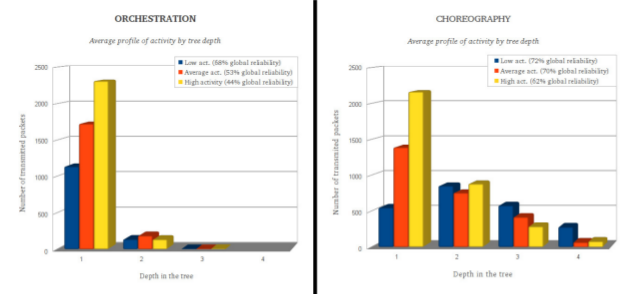} 
\caption{In an \textit{Orchestration}, top level nodes are heavily used because all data must travel to the sink, and come back to actuators. In a \textit{Choreography}, some messages reach their destination more directly, reducing the burden of upper nodes.} 
\label{fig:ORCHChorrouting} 
\end{figure} 

The impact of the \textit{Orchestrated} architecture of a software over the RPL network is displayed on the first graph (Figure~\ref{fig:ORCHChorrouting} left). The average number of forwarded packets is given depending on the node level in the tree. The first conclusion is that the traffic is mainly concentrated on top level nodes. Because of its organization, an \textit{Orchestrated} software implies the transmission through upper level nodes. When the global activity is important, the top level nodes reach saturation. For the higher activity class, we can see that the activity decreases on lower levels. There is a bottleneck at the top of the tree, and messages are lost. The global reliability of the network decreases quickly. In comparison, the deeper in the tree a node is, the less it is requested. At level 3, traffic becomes insignificant because messages are lost at upper levels. In this \textit{Orchestrated} design, all application messages go up to the sink. As a result, the whole network reliability is highly correlated with the overloading of the tree top-level nodes. The more these nodes are overloaded, the less reliable the network is.

On the right of Figure~\ref{fig:ORCHChorrouting}, the profile of a \textit{Choreographed} software reveals to be different. The global activity is better distributed all over the network. Some messages find their way at lower levels. In general, these levels handle more messages than in an \textit{Orchestration} because some paths avoid top-level nodes. While traffic decreases on top-level, levels 2, 3 and 4 forward more packets. Some of them have their origin and destination at these low levels, leveraging upper levels. The network reliability increases, even in the heaviest of our 3 activity classes.

\subsection{Testbed's global view: activity and network reliability}
\begin{figure} 
\centering 
\includegraphics[width=12cm]{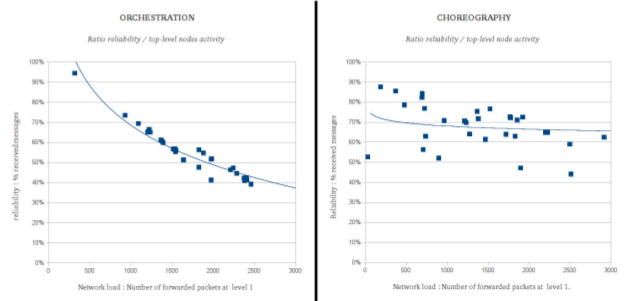} 
\caption{The more first level is overload, the less reliable is the network. \textit{Orchestration} saturates top level nodes. In a \textit{Choreography}, there is no correlation between reliability and first level activity. The network traffic increases, more
depending on messages destination rather than their amount.} 
\label{fig:ORCHCHORratio} 
\end{figure} 
The second information given by this experiment comes from the raw results (before average) of the packet delivery ratio (PDR). We were trying to see if a link appears between the application architecture and the overall reliability of the network. We wanted to see if the choice of an architecture for an application has a strong impact on how the load is spread over the network and the global reliability.

To search for a correlation, we have plotted the top level activity PDR Figure~\ref{fig:ORCHCHORratio}. Because of the data amount in our experiment, the network is stressed, leading to significant packet losses. In the case of \textit{Orchestration}, the link between top level nodes activity and network reliability is clear. The transmission limit of that level defines the entire network accessibility. Figure~\ref{fig:ORCHCHORratio} (Orchestration is on the left) shows the global trend.

On the right of Figure~\ref{fig:ORCHCHORratio}, the \textit{Choreography} organization does not reveal link between these two metrics. Traffic is better distributed within the network. Some exchanges don't travel across the top of the tree. The global reliability reaches higher values, while being less correlated to top level nodes activity.

Figure~\ref{fig:ORCHCHORratio} shows the impact of application's design over a constrained network. The choice of an \textit{Orchestrated} architecture decreases the global reliability when the traffic increases. Figure~\ref{fig:ORCHChorrouting} shows how the top level nodes are affected by this choice, leading to a higher transmissions activity. With \textit{Orchestrated} architecture, some top level nodes may run quickly out of energy, eventually forcing the choice of another path (by rebuilding the tree). The most requested nodes (those making the link with the sink) are then overloaded. In contrast, the same application designed as a \textit{Choreography} (when it is possible) spreads the activity most randomly all over the network. Various nodes are involved, the network is more reliable and has a longer lifetime.

\section{D-LITe}
Designing a Choreography involves the ability to program each stakeholder, in order to create direct interactions. To achieve a usable IoT development framework, we have summarized the needs and constraints of applications in this domain. Our proposal is based on a universal network access and programming approach of each component. The main trend in the IoT field is to propose a common protocol for describing objects capability. Our solution proposes to allow the universal programming of Objects, regardless its specificities. In D-LITe, Objects are accessible through REST methods, and can be programmed in an easy way, by the use of \textit{Finite State Automaton}. An application is built as a Choreography of Services, as part of Services Oriented Application (SOA).
\subsection{REST or CoAP access}
SOA introduces loose coupling between applications and services while providing hardware independence. Many protocols realize SOA. One of them is SOAP~\cite{SOAPw3c} which is adapted to the Internet of Data, where all stakeholder are powerful and not limited in energy. But SOAP is too verbose for Sensors Networks. Their constraints in terms of payload and network throughput limit the use of an XML parser.

REST architecture~\cite{RESTFielding} is an alternative that uses standard HTTP methods. REST is lightweight and simple to adapt to our purpose. As even REST is too heavy for very constrained Object, CoAP~\cite{shelby2010embedded} \emph{``extends the REST architecture to a suitable form for the most constrained nodes''} of WSAN~\cite{shelby2010constrained}. CoAP is build over 6LowPAN, an IPv6 version adapted to WSAN~\cite{Dunkels2004makingIPforWSN}\cite{Shelby20106lowpan}. By implementing HTTP over UDP, and using compression of HTTP methods, CoAP is designed to simply permit translations between the standard and universal REST commands from the Internet and a 6LowPAN Network, while being particularly suitable to the limited payload of smart objects networks. It is also easily adapted to usual REST network, leading to quick translations from and to the Internet of PC.

\subsection{Hardware Abstraction layer}
\begin{figure} 
\centering         
\includegraphics[width=10cm]{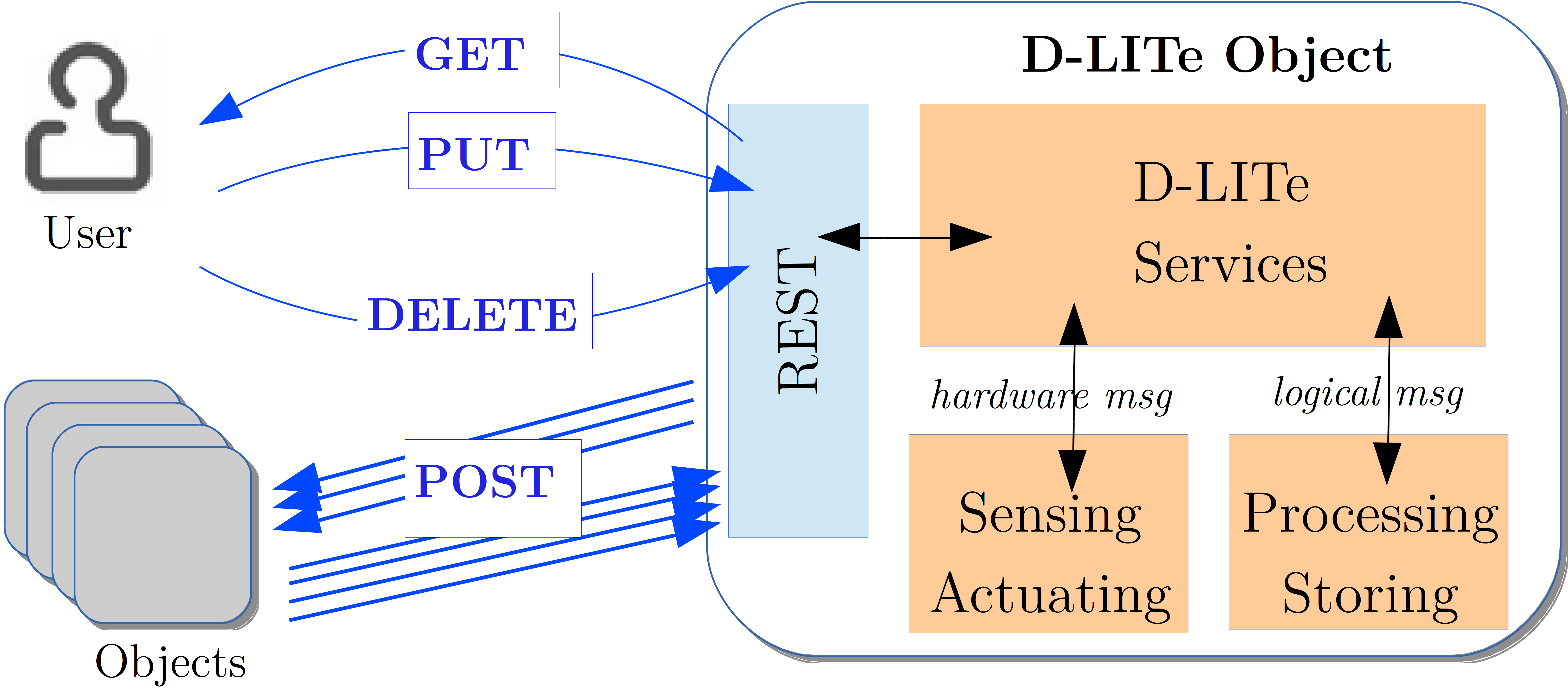} 
\caption{D-LITe architecture overview. After receiving user's description of the application logic, a \textit{D-LITe} Object reacts and generates external, hardware (sensing/actuating) or logical messages (testing/creating variables). These services are accessible through REST methods.}
\label{fig:Architecture} 
\end{figure} 
IoT developers need a standardized platform in order to create or modify IoT applications quickly and easily. These applications need to be promptly deployable at any time, ``over the air'' if possible. They are user specific, depending on user's objects and their configurations, and needs. The customization of user's applications involves flexibility.

D-LITe \textit{(Distributed Logic for Internet of Things sErvices)}~\cite{cherrier2011Dlite} introduces a hardware abstraction to get rid of \textit{language/operating system/technical} constraints. D-LITe framework hides tricky accesses to hardware functionalities by offering a simplified and logical view of its features. Our platform offers standard ways to access these features through the use of plain text messages. These messages (that the developer needs to know) are presented during node's discovery phase. They can be used by the programmer in his logical description (his software). They give access to Object's functionalities, depending on the message's type (external, hardware, or logical). In fact, one can say that \textit{external messages} are for the outside world (useful to make Objects interact) while \textit{hardware} and \textit{logical} messages are for Object's internal use (sensing, actuating, or computing operations offered by the Object) (Fig~\ref{fig:Architecture}).

\subsection{D-LITe usage of REST}
In our ``Object-as-a-service'' approach, we need to hide Object's specificities and to offer some management facilities. The composition of Objects in an IoT application point-of-view requires that each Object functionalities are discovered, and then used in order to participate to the global interactions needed by the user. An Object must be able to describe itself, to be programmed, to compute and react to its environment, to exchange the result of its treatments with other Objects.

The 4 REST commands (\textit{GET}, \textit{PUT}, \textit{POST} and \textit{DELETE}, Figure~\ref{fig:Architecture}) are used by D-LITe to achieve these goals:
\begin{itemize}
  \item \textbf{GET} gives the description of the targeted Object. Mainly, it contains the D-LITe platform version and the supported features list. Each feature describes the hardware that this Object (more precisely the D-LITe virtual machine) can deal with.
  \item \textbf{PUT} offers the ability to dynamically program this Object. A list of instructions, describing its role in the whole application, is sent to the Object. This list of instructions, called \textit{Behaviour}, is the program to be executed by the node. It has inputs (from other nodes, or the hardware of the Objects), and outputs (outgoing messages, for the hardware of the object, or other nodes.) A subscribers list is also given in the PUT message. Each Object of the list will receive all the outgoing messages.
  \item \textbf{POST} transfers messages to subscribers. Each Object of the subscribers list (given when programming this Object with \textit{PUT}command) will received the outgoing messages (the output of the program described in the \textit{PUT} command).
  \item \textbf{DELETE} clears the Object \textit{Behaviour} and subscribers list, and is called before the reprogramming of the Object (before a \textit{PUT}).
\end{itemize}

\subsection{Implementations}
\begin{figure} 
\centering         
\includegraphics[width=8cm]{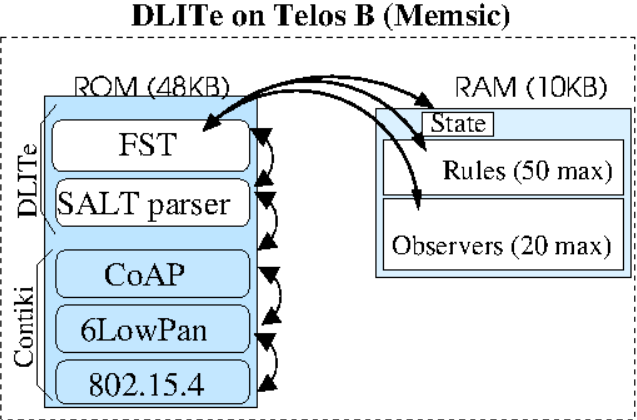} 
\caption{DLITe implementation with Contiki on TelosB.} 
\label{fig:TelosB} 
\end{figure} 
First version of DLITE was written for Contiki (Feb,17 2011 daily snapshot) and tested on some TelosB (http://advanticsys.com). The actual binary size of D-LITe is 47KB and fits in the 48KB program flash memory of a TelosB (Figure~\ref{fig:TelosB}). Programming TelosB is usually done by flashing the ROM through USB connector. According to our vision, the software is divided in two parts. The first one is the D-LITe framework (the fixed part) and the other is the user's logic description (the variable part). The node is flashed once (for the installation of the fixed part, D-LITe). Then, on-the-fly, developers send through the network, at any time, the specific logical description of the node's behaviour (the variable part). They also give the list of subscribers (others Objects that are related to this one). All that dynamic part is stored in RAM (10 KB for the TelosB). The D-LITe framework then executes the logic. All manipulated data are 6 characters long. A rule's length (an element of the program) is 24 Bytes (4 words of 6 characters, 2 states and 2 messages). Our current implementation can handle up to 50 rules. A list of 20 Observers maximum is stored in 16 Bytes each (IPv6 of subscribers). 

A second version of D-LITe is available for Android smartphones and Tablets. Because these Objects are more powerful, they offer a wide range of sensing and actuating capabilities. We also need some visual effects and interactions on these devices. Our application is bigger (1,93 MB)\footnote{accessible here: \url{http://igm.univ-mlv.fr/~cherrier/download/Android-DLite.apk}}, and can send SMS, use the accelerometer, display virtual button, light, etc. We also have a virtual Object coded in Java, that is used to simulate nodes in wider applications.

\section{SALT}
D-LITe is the software we propose to install on each Object in order to build IoT applications. This platform hides object's specificities, and presents a universal view of the computing capabilites of this object. D-LITe uses a specific language called SALT (Simple Application Logic description using Transducers)\cite{cherrier2013SALT} to describe the logic to be executed on nodes. SALT is based on \textit{Transducers}, with some specific extensions provided for the IoT. We chose Transducers because they are well-known, because the parser is easy to write, has small memory footprint, and a small CPU load when interpreting the rules. 

\subsection{Finite State Transducers}

SALT expresses an algorithm as a \textit{Finite State Transducer} (FST). FST are \textit{Finite State Automata} with an additional output Alphabet. Its formal representation is a 6-tuple $T(Q,\Sigma,\Gamma,I,F,\delta)$. SALT defines the meaning of each element as follows:
\begin{itemize}
 \item $Q$: \emph{States} supported by the FST,
 \item $\Sigma$: \emph{Input Messages} the node reacts to,
 \item $\Gamma$: \emph{Output Messages} that can be sent,
 \item $I$: \emph{Initial State} (only one in D-LITe),
 \item $F$: \emph{Final States} (often none in IoT application),
 \item $\delta$: Transitions between States (the distributed ``\emph{logic}'').
\end{itemize}
SALT uses \emph{Input Messages} and \emph{Output Messages} instead of alphabets. These messages are used to exchange stimuli between Objects. They also give access to hardware through the \textit{Hardware Abstraction Layer} provided by D-LITe (Fig~\ref{fig:graphical}). As seen above, this abstraction layer hides the real Object, and can be invoked by the meaning of simple words of the alphabet. These simple words abstract the complex programming calls. The compact Transducer's description allows the transmission of the whole program through the network using the \textit{REST deployment service (PUT)} provided by D-LITe. The targeted Object's behaviour can be set dynamically without any physical access. 

The universality given by \textit{FST} and \textit{Hardware Abstraction Layer} helps to build IoT applications. The expressibility of Transducers needs to be extended to build Choreographed applications as powerful as Orchestrated ones. With its alphabet extensions, SALT gives an access to data management and timers in respect of FST philosophy.

\subsection{Message Extensions}

\begin{table}
\caption{Choreography issues, and SALT proposals} 
\begin{tabular}{|p{3 cm}|p{9 cm}|}
\hline
\textit{Describing the logic}&SALT can describe a Transducer understood by Objects\\
\hline
\textit{Objects cooperating}& Transducer's output is sent to other Object and becomes their input. Objects react in cascade.\\
\hline
\textit{Updating the logic}&SALT description of the logic has a very limited size, because it uses the framework installed on each Object. The logic can be changed, quickly, without any physical access and in few network messages.\\
\hline
\textit{Hardware agnostic}&SALT uses the hardware abstraction layer provide by D-LITe. The Input or Output alphabet of the Transducer gives accesses to hardware specificities.\\
\hline
\textit{Computation and timers}&Improvement for IoT application is obtained by adding time control and variables management. Alphabets use \textit{extensions} to offer a greater level of expressibility and autonomy. \\
\hline
\end{tabular}
\label{table:SALTProposal} 
\end{table}

Choreographies move the application logic from the central control point to Objects themselves. In order to be nearly as powerful as Orchestrations, IoT Choreographies must be able to:
\begin{itemize}
 \item  \textbf{store}, \textbf{test} and \textbf{compute} values. An Object can inferred more elaborated mechanisms from its own measures, needed for its autonomy.
 \item limit reactions over \textbf{time} as part of the algorithm (for example counting replies during a certain period of time).
\end{itemize}

X.~Fu et al in \cite{fu2004model} introduce the use of ``\textit{guarded automata}'' for Web Services Choreography. A guarded automata pays attention to the message content and not only its concordance to the given reference string. We extend this idea to increase the semantic of the exchange. Unlike simple words of an alphabet, a SALT message has an inner meaning. This allows the test of variables content, arguments management, and mathematical operations expressions. A message such as ``$=(count,1)$'' stands for \textit{store the value 1 in a variable named ``count''}. ``$<(i,3)$'' means \textit{if the variable i is less than 3}.

Extended messages give SALT timers, variables management and improve FST expressibility. Most of Orchestrated application control structures made at the top-level can be achieved in a Choreography, at end-points, according to its specific distributed organization (Table~\ref{table:SALTProposal}).

Extended messages, allied with the event-centric approach of our solution, make it easier to build distributed IoT applications, as they offer a simple programming language adapted to the needs of IoT applications. Programs made with SALT can be sent to the Object over the air, in a small footprint adapted for the constrained Objects and network such as WSAN. 

\subsection{Using SALT}
In SALT, extended \emph{Input Alphabet} and \emph{Output Alphabet} give semantic to these messages. As each Object receives its Transducer (Fig~\ref{fig:Architecture}), the embedded framework D-LITe executes it and links these logical expressions to actions offered by Object's hardware. For SALT, a Transducer always starts with a unique \textit{Initial State}. Then, received messages are checked as \textit{Transducer Input}, and the matching \textit{Transition} leads to the next \textit{State}. The \textit{Transducer} sends the \textit{Output Message} given in the \textit{Transition}. 4 items are given in a \textit{Transition}: The \textit{From State}, the \textit{Received Message}, the \textit{Output Message}, and the \textit{Next State}.

\textit{Input} and \textit{Output} alphabets describe how to sense or to actuate the real world, do basic processing operations, or exchange with other Objects of an IoT application. These messages are of 3 different kinds:
\begin{itemize}
 \item \textit{External messages} organize the collaboration between Objects. Messages content becomes the input of the Transducers running on each listening Object (of the subscribers list)
 \item \textit{Hardware messages} link the developer's logic to the Object's hardware. This list of specific messages is defined during discovery phase:
      \begin{enumerate}
      \item Sensing generates \textit{Input} messages (e.g. a button has been pushed, the temperature has changed),
      \item Transducer's \textit{Output} uses Object's actuating capabilities (e.g. switching on/off),
      \item Timers are interpreted as virtual sensing and actuating capabilities. An \textit{Output} hardware message starts the timer and defines the delay. The Transducer receives an \textit{Input} message when the timer expires.
  \end{enumerate}
 \item \textit{logical messages} define, alter and test variables. Logical \textit{Output} messages are used to create a variable and/or alter its content. On the contrary, variables are tested with an \textit{Input} message. If the test is true, the \textit{Transition} is executed as if an event has occurred.
\end{itemize}

\begin{figure}[t] 
\centering         
\includegraphics[width=8cm]{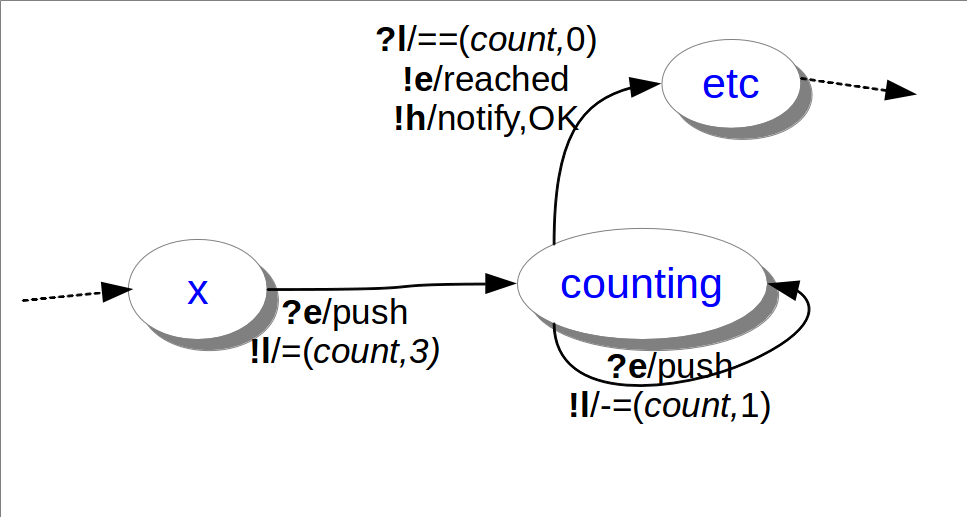} 
\caption{SALT Graphical representation, for counting 4 ``push'' messages then send ``reached'' and display ``OK'' on the device. The first message creates a variable named \textit{count} and the following decrement it. 0 is detected by the last transition.}
\label{fig:graphical} 
\end{figure} 
\begin{table}
\caption{SALT textual  representation of Example
 Fig~\protect\ref{fig:graphical}} 
\begin{tabular}{|p{12 cm}|}
\hline
\textit{x}  ?\textbf{e}/push  !\textbf{l}/=(count,3)  \textit{counting}\\
\textit{counting} ?\textbf{e}/push  !\textbf{l}/-=(count,1) \textit{counting}\\
\textit{counting} ?\textbf{l}/==(count,0) !\textbf{e}/reached !\textbf{h}/notify,OK \textit{etc}\\
\hline
\end{tabular}
\label{table:SALTExemple} 
\end{table}

In other words, external messages are intended for Object's environment \textit{communication} (a developer uses them to make Objects react) while hardware and logical messages are for \textit{hardware internal use} (sensing, actuating, or computing) (Fig~\ref{fig:Architecture}).

\subsection{Links to the Hardware Abstraction Layer}

\begin{table}
\caption{correspondence between features and Input-Output Alphabets (extract) }
\begin{tabular}{|c|c| p{3.4cm} | p{2.8cm} |}
\hline
\multicolumn{4}{|c|}{\textbf{Sensors}} \\
\hline
Feature & Abstraction & Messages & arguments \\
\hline
  \textit{button} & Button, switch & push & \\
\hline
  \textit{acceleration} & Movement sensor &  accelerationChanged & X,Y,Z\\
\hline
  \textit{gps} & GPS & positionChanged & long.,lat.,alt. \\
\hline
  \textit{joystick} & Joystick & Click \newline RightUp \newline CenterDOWN \newline Right ... &  \\
\hline
  \textit{light} & lightning sensor & lightChanged & X \\
\hline

\multicolumn{4}{|c|}{\textbf{Actuators}} \\
\hline
Feature & Abstraction & Messages & arguments \\
\hline
  \textit{led} &  led, lights, etc. & led  & on \newline off \\
\hline
  \textit{notification} &  Displayer, screen, etc. & notify  & String \\
\hline
  \textit{sms} &  Telephone & sms  & number,message \\
\hline
\end{tabular}
\label{table:Features} 
\end{table}

As it gives a universal view of the Object, D-LITe has to offer an access to each Object functionalities. The SALT language makes the use of this D-LITe Object through specific words of the Transducer's Alphabet. The Hardware Abstraction Layer implemented in D-LITe make the link between these words and the functions offered. Depending on Object's type (that we call \textit{Features}), the words list is defined (see Table~\ref{table:Features}). For actuators, these words (used as Output) will trigger the intended action. For example, a D-LITe lamp will switch on when its Transducer sends the word ``\textit{led on}''. On the other side, the hardware of a sensor sends a referenced word in the Transducer's Input, so a transition can be executed. For example, the HAL receives ``\textit{push}'' as an input message when a user presses the button of an Object with such a feature. If this Object is a GPS, the HAL will receive ``\textit{positionChanged}'' followed by coordinates when it has moved.

\begin{figure} 
\centering         
\includegraphics[width=9cm]{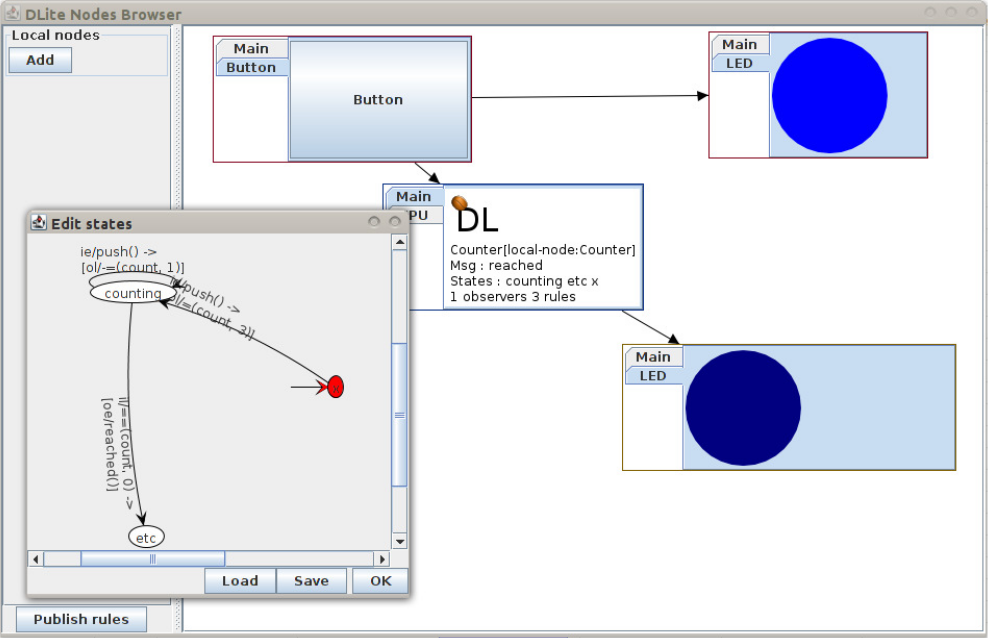} 
\caption{SALT GUI tool. On this example, a button (top left) controls a light (top right), but is also observed by an indicator led (bottom right) running a counter (counting 3 pushes). The Transducer (the program that counts the 3 pushes) is displayed near its targeted Object. Each Object has its own Transducer (The button sends push, and the top led waits for them). The ``publish rules'' button deploys the whole application.}
\label{fig:SaltGUI} 
\end{figure}

The Table~\ref{table:Features} shows some properties of the Features words list. The Transducer given in Figure~\ref{fig:graphical} is able to display a message on a screen, as it uses the hardware word ``\textit{notify}'' (\textbf{h} tells D-LITe that the message is for the Hardware). The argument ``\textit{OK}'' is displayed because the Hardware Abstraction Layer implements this call. \textit{Features} are detected during Object's discovery (sent in the response to the GET request). The programming tool~\ref{fig:SaltGUI} checks that the programmer uses only words supported by the hardware, as indicated in Table~\ref{table:Features}.

\subsection{Language description}
This section gives an overview of the formalism used in SALT to fully described the Transducers used in IoT applications.

SALT expresses the complete description of the Transducers by its \textit{Transitions} only. The formalism of a \textit{Transition} description is given by Table~\ref{table:SALTTransition}. The \textit{States} represent steps in the algorithm, and their names are content-free strings. \textit{Input} and \textit{Output} alphabets have a more constrained formalism. 

There is usually one Input message in a Transition ($\epsilon$, the empty message, can be used as input. The transition is therefore immediately executed. It is useful to set variables or to configure the Object's hardware). A transition has zero to many Output Messages:
\begin{itemize}
 \item no Output message to move to the next FST's state without altering other Object or making specific change inside the node (variable, actuating),
 \item one or many messages describing actions to do or/and messages to be sent to listeners.
\end{itemize}
The message type is coded with one char (\textbf{e})xternal, (\textbf{l})ogical, or (\textbf{h})ardware, and preceded by a question mark (input message) or an exclamation mark (output message). The message contains a simple word to be sent to listeners, or timer's duration, variable's value or test, depending on its type.
\begin{table}[t]
\caption{SALT language: Transition structure} 
\begin{tabular}{|p{1.7 cm}|p{10.3 cm}|}
\hline
\multicolumn{2}{|c|} {Transition Formalisation : \textbf{State   Input  Output\ldots State}}\\
\hline
State\newline \textbf{xxxx}&A single word \textbf{xxxx}, giving semantic to programmer's reasoning, having no effect on anything. It's only a name for the current stage in the Object's logic.\\
\hline
Input\newline \textbf{?x/yyyy}& a \textbf{question mark}, followed by (x) (\textbf{e}, \textbf{l} or \textbf{h} for external, logical or hardware message). Extended message (\textbf{yyyy}) has different formalism. For \textit{external}, it is the expected content. For \textit{logical}, \textbf{yyyy} is about a variable (for example, testing its content). Finally, \textbf{yyyy} in \textit{hardware} message is a pre-defined command using Object's sensing capabilities (described during Object's discovery phase).\\
\hline
Output\newline \textbf{!x/yyyy}& a \textbf{exclamation mark}, followed by (x) (\textbf{e}, \textbf{l} or \textbf{h}). An \textit{external} message is sent to all listeners, a \textit{logical} message creates or alters a variable inside the Object, while an \textit{hardware} message actuates something. This action can be parametrized if needed, for example to set the hardware (i.e. \textit{led(on)} or \textit{led(off)}) or for processing operation (i.e \textit{=(nb,5)} or \textit{+=(count,2)}).\\
\hline
\textbf{\ldots}& A transition may have multiple output messages (for Object's hardware, listeners or variable management)\\
\hline
State\newline  \textbf{zzzz}&A simple word, describing the step reached by the application in developer's mind.\\
\hline
\end{tabular}
\label{table:SALTTransition} 
\end{table}

\begin{itemize}
  \item \textbf{External} messages contain a string (received in the case of \textit{input} messages, or sent for \textit{output} messages), i.e. ``on'', ``cold'', ``intrusion'', ``fire''.
  \item \textbf{Hardware} messages use keywords related to the event being managed. Keywords list is defined in Table~\ref{table:Features} depending of the kind of Object, and hard coded in D-LITe. Used as \textit{Input messages}, these keywords match sensed events. In \textit{Output messages}, they trigger actions. Some of these \textit{Input/Ouput} may have parameters.
  \item \textbf{Logical} messages are for variable management and content testing. They use operation signs and variable names (Table~\ref{table:SALTLogical}). A developer sets or modifies a variable in \textit{Output messages} with $=$,$+=$,$-=$,$*=$ or $/=$(Fig~\ref{fig:graphical}). In \textit{Input messages}, the added semantic is used to test variables content, with \textit{$==$}, \textit{$!=$},\textit{$>$},\textit{$<$},\textit{$>=$} or \textit{$<=$}.
\end{itemize}

\begin{table}
\caption{SALT language: Extensions to express variable management} 
\begin{tabular}{|p{3 cm}|p{8.5 cm}|}
\hline
\multicolumn{2}{|c|}{\textbf{Input} extended message
(\textit{testing variables})}\\
\hline
\textbf{$?l/==$(var,value)}&If ``\textit{var}'' is equal to ``\textit{value}''. If true, the transition using \textbf{$<$} \textbf{$>$} \textbf{$<=$} \textbf{$>=$} and \textbf{$!=$} is executed\\
\hline
\multicolumn{2}{|c|}{\textbf{Output} extended message (\textit{setting or computing variables})}\\
\hline
\textbf{$!l/=$(var,value)}&Set ``\textit{var}'' to ``\textit{value}''. The variable is created if unknown.\\
\hline
\textbf{$!l/+=$(var,value)}&Add ``\textit{value}'' to ``\textit{var}''. \textit{Value} can be another variable. The result is stored in ``\textit{var}''. 4 operations are supported (\textbf{$+=$},\textbf{$-=$}, \textbf{$*=$} and \textbf{$/=$})\\
\hline
\end{tabular}
\label{table:SALTLogical} 
\end{table}

\section{Conclusion}
In this paper, we have described our proposition to realize real \textit{Internet of Things} applications. The main idea stands in organizing a Choreography of virtual representations of each Object involved in the application. Instead of defining a common description for each type of objects to build centralized application, we propose to access the programming ability of each Object, regardless its type. Each Object of an IoT application is able to compute and to communicate through the network. Our solution capitalizes on these two characteristics and takes into account Object and network constraints. Our framework is lightweight, the programming language is simple, but able to build a wide range of applications.

First, we have studied the impacts of our proposed architecture over WSAN, one of the main constrained network involved in IoT domain. A distributed organization has already been described as more adapted for these networks, but we have quantified the gain. A distributed architecture (a Choreography) running over a network organized as a tree, reduces up to 66\% the average path length between nodes. Trees are the form used in the two main solution we are working on (ZigBee and 6LowPAN). In IoT application, nodes interact with each others, giving a Choreographed architecture a true added value, especially in non-reliable and energy constrained networks. 

The next part of the paper is devoted to D-LITe which aims to build Choreographies. D-LITe offers the virtual representation and a universal network access to each Object. This access is realized following a REST architecture. REST is used to implement a service to discover Object characteristics, and to access the programming capabilities of the Object. This service is used to deploy the logic the Object has to follow (as part of the IoT application). By introducing a hardware abstraction layer, D-LITe hides the complex calls to functionalities offered by Objects (sensing capabilities in the case of a sensor, actuating commands for an actuator, and other actions a virtual Object can provide, such as sending an SMS, gathering data from a web server, publishing a message for a Instant messaging service). D-LITe can be programmed in SALT, a specific language created to match the needs of IoT application while taking into account the constraints of the small devices involved. SALT is derived from Finite State Transducers, with several extensions. It uses extended messages to add expressibility to the exchanges (between nodes, or between the hardware and our framework). Compositions of logical units (Choreographies) written in SALT reach the expressiveness of Orchestrations. 

In the future, we will focus on a higher vision of the IoT application, to reach better interactions between Objects and to improve the dynamic integration of new Objects in a running application.

\bibliographystyle{abbrv}
\bibliography{DLITE-SALT}

\end{document}